

\font\titlefont = cmr10 scaled\magstep 4
\font\sectionfont = cmr10
\font\littlefont = cmr5
\font\eightrm = cmr8

\def\ss{\scriptstyle}
\def\sss{\scriptscriptstyle}

\newcount\tcflag
\tcflag = 0  


\ifnum\tcflag = 0 \magnification = 1200 \fi  

\global\baselineskip = 1.2\baselineskip
\global\parskip = 4pt plus 0.3pt
\global\abovedisplayskip = 18pt plus3pt minus9pt
\global\belowdisplayskip = 18pt plus3pt minus9pt
\global\abovedisplayshortskip = 6pt plus3pt
\global\belowdisplayshortskip = 6pt plus3pt

\def\barsoff{\overfullrule=0pt}


\def\endignore{}
\def\ignore #1\endignore{}

\newcount\dflag
\dflag = 0


\def\monthname{\ifcase\month
\or January \or February \or March \or April \or May \or June%
\or July \or August \or September \or October \or November %
\or December
\fi}

\newcount\dummy
\newcount\minute  
\newcount\hour
\newcount\localtime
\newcount\localday
\localtime = \time
\localday = \day

\def\advanceclock#1#2{ 
\dummy = #1
\multiply\dummy by 60
\advance\dummy by #2
\advance\localtime by \dummy
\ifnum\localtime > 1440
\advance\localtime by -1440
\advance\localday by 1
\fi}

\def\settime{{\dummy = \localtime%
\divide\dummy by 60%
\hour = \dummy
\minute = \localtime%
\multiply\dummy by 60%
\advance\minute by -\dummy
\ifnum\minute < 10
\xdef\spacer{0} 
\else \xdef\spacer{}
\fi %
\ifnum\hour < 12
\xdef\ampm{a.m.} 
\else
\xdef\ampm{p.m.} 
\advance\hour by -12 %
\fi %
\ifnum\hour = 0 \hour = 12 \fi
\xdef\timestring{\number\hour : \spacer \number\minute%
\thinspace \ampm}}}



\def\endtitle{}
\def\title#1\endtitle{\vskip.5in\titlefont
\global\baselineskip = 2\baselineskip
#1\vskip.4in
\baselineskip = 0.5\baselineskip\rm}

\def\endauthors{}
\def\authors#1\endauthors{#1}

\def\endabstract{}
\def\abstract#1\endabstract{\vskip .3in%
\centerline{\sectionfont\bf Abstract}%
\vskip .1in
\noindent#1}

\def\nopageonenumber{\footline={\ifnum\pageno<2\hfil\else
\hss\tenrm\folio\hss\fi}}  

\newcount\nsection
\newcount\nsubsection

\def\section#1{\global\advance\nsection by 1
\nsubsection=0
\bigskip\noindent\centerline{\sectionfont \bf \number\nsection.\ #1}
\bigskip\rm\nobreak}

\def\subsection#1{\global\advance\nsubsection by 1
\bigskip\noindent\sectionfont \sl \number\nsection.\number\nsubsection)\
#1\bigskip\rm\nobreak}

\def\topic#1{{\medskip\noindent $\bullet$ \it #1:}}
\def\endtopic{\medskip}

\def\appendix#1#2{\bigskip\noindent%
\centerline{\sectionfont \bf Appendix #1.\ #2}
\bigskip\rm\nobreak}


\newcount\nref
\global\nref = 1

\def\ref#1#2{\xdef #1{[\number\nref]}
\ifnum\nref = 1\global\xdef\therefs{\item{[\number\nref]} #2\ }
\else
\global\xdef\oldrefs{\therefs}
\global\xdef\therefs{\oldrefs\vskip.1in\item{[\number\nref]} #2\ }%
\fi%
\global\advance\nref by 1
}

\def\listrefs{\vfill\eject\section{References}\therefs}


\newcount\nfoot
\global\nfoot = 1

\def\foot#1#2{\xdef #1{(\number\nfoot)}
\footnote{${}^{\number\nfoot}$}{\eightrm #2}
\global\advance\nfoot by 1
}


\newcount\nfig
\global\nfig = 1

\def\fig#1{\xdef #1{(\number\nfig)}
\global\advance\nfig by 1
}


\newcount\cflag
\newcount\nequation
\global\nequation = 1
\def\eqlabel{(1)}

\def\nexteqno{\ifnum\cflag = 0
\global\advance\nequation by 1
\fi
\global\cflag = 0
\xdef\eqlabel{(\number\nequation)}}

\def\lasteqno{\global\advance\nequation by -1
\xdef\eqlabel{(\number\nequation)}}

\def\label#1{\xdef #1{(\number\nequation)}
\ifnum\dflag = 1
{\escapechar = -1
\xdef\draftname{\littlefont\string#1}}
\fi}

\def\clabel#1#2{\xdef\eqlabel{(\number\nequation #2)}
label
\global\cflag = 1
\xdef #1{\eqlabel}
\ifnum\dflag = 1
{\escapechar = -1
\xdef\draftname{\string#1}}
\fi}

\def\cclabel#1#2{\xdef\eqlabel{#2)}
\global\cflag = 1
\xdef #1{\eqlabel}
\ifnum\dflag = 1
{\escapechar = -1
\xdef\draftname{\string#1}}
\fi}


\def\eeq{}

\def\eqnn #1\eeq{$$ #1 $$}

\def\eq #1\eeq{
\ifnum\dflag = 0
{\xdef\draftname{\ }}
\fi 
$$ #1
\eqno{\eqlabel \rlap{\ \draftname}} $$
\nexteqno}







\def\eqa #1\eeq{
\ifnum\dflag = 0
{\xdef\draftname{\ }}
\fi 
$$ \eqalignno{ #1 } $$
\global\cflag = 0}


\def\ie{{\it i.e.\/}}
\def\eg{{\it e.g.\/}}
\def\etc{{\it etc.\/}}

\def\apriori{{\it a priori\/}}

\def\via{{\it via\/}}


\def\anp#1#2#3{{\it Ann.\ Phys. (NY)} {\bf #1} (19#2) #3}

\def\cmp#1#2#3{{\it Comm.\ Math.\ Phys.} {\bf #1} (19#2) #3}
\def\ijmp#1#2#3{{\it Int.\ J.\ Mod.\ Phys.} {\bf A#1} (19#2) #3}

\def\npb#1#2#3{{\it Nucl.\ Phys.} {\bf B#1} (19#2) #3}
\def\plb#1#2#3{{\it Phys.\ Lett.} {\bf #1B} (19#2) #3}

\def\prb#1#2#3{{\it Phys.\ Rev.} {\bf B#1} (19#2) #3}

\def\prd#1#2#3{{\it Phys.\ Rev.} {\bf D#1} (19#2) #3}
\def\pr#1#2#3{{\it Phys.\ Rev.} {\bf #1} (19#2) #3}
\def\prep#1#2#3{{\it Phys.\ Rep.} {\bf C#1} (19#2) #3}


\global\nulldelimiterspace = 0pt



\def\frac#1#2{{{#1} \over {#2}}\,}  
\def\hf{{1\over 2}}
\def\nth#1{{1\over #1}}



\def\Square{{\vbox
{\hrule height 0.6pt\hbox{\vrule width 0.6pt\hskip 3pt

\vbox{\vskip 6pt}\hskip 3pt \vrule width 0.6pt}\hrule height 0.6pt}}}
\def\Dsl{\hbox{/\kern-.6700em\it D}} 
\def\dsl{\hbox{/\kern-.5300em$\partial$}}
\def\pxpsl{\hbox{/\kern-.5600em$p$}}
\def\ssl{\hbox{/\kern-.5300em$s$}}
\def\epssl{\hbox{/\kern-.5100em$\epsilon$}}
\def\delsl{\hbox{/\kern-.6300em$\nabla$}}
\def\lxpsl{\hbox{/\kern-.4300em$l$}}
\def\elxpsl{\hbox{/\kern-.4500em$\ell$}}
\def\kxpsl{\hbox{/\kern-.5100em$k$}}
\def\qxpsl{\hbox{/\kern-.5000em$q$}}
\def\sla#1{\raise.15ex\hbox{$/$}\kern-.57em #1}
\def\Pl{\gamma_{\sss L}}
\def\Pr{\gamma_{\sss R}}



\def\roughly#1
{\mathrel{\raise.3ex\hbox{$#1$\kern-.75em\lower1ex\hbox{$\sim$}}}}

\def\tw#1{\tilde{#1}}
\def\ol#1{\overline{#1}}





\def\Scd{{\cal D}}

\def\Scl{{\cal L}}

\def\Sco{{\cal O}}


\def\ssa{{\sss A}}
\def\ssb{{\sss B}}

\def\ssf{{\sss F}}
\def\ssg{{\sss G}}

\def\ssl{{\sss L}}

\def\ssp{{\sss P}}

\def\ssr{{\sss R}}


\def\diag#1{{\rm diag}\left( #1 \right)}



\def\avg#1{\langle #1 \rangle}



\def\hc{{\rm h.c.}}




\nopageonenumber
\baselineskip = 18pt
\barsoff



\def\veps{\varepsilon}

\def\IR{\relax{\rm I\kern-.18em R}}
\def\psibar{\ol{\psi}}

\def\lft{\ssl}
\def\rht{\ssr}
\def\bk{\item{}}
\def\twL{\tw\Lambda}


\rightline{December, 1993.}
\rightline{McGill-93/45, NEIP-93-010}
\rightline{hep-th/9401105}
\vskip .6in

\centerline{$\hbox{\titlefont Bosonization as Duality}^\dagger
$\footnote{}{${}^\dagger$
\eightrm Research supported by the Swiss National Foundation.}}
\bigskip

\vskip 0.25in
\authors
\centerline{C.P. Burgess${}^*$\footnote{}{${}^*$
{\eightrm Permanent Address:
Physics Department, McGill University,
3600 University St., Montr\'eal,
}}\footnote{}{\eightrm $\phantom{{}^*}$ Qu\'ebec,
 Canada, H3A 2T8. E-mail: cliff@physics.mcgill.ca.}
and F. Quevedo${}^{**}$\footnote{}{${}^{**}$ {\eightrm E-mail:
quevedo@iph.unine.ch.}}}
\vskip .25in
\centerline{\it Institut de Physique}
\vskip 0.05in
\centerline{\it Universit\'e de Neuch\^atel}
\vskip 0.05in
\centerline{\it CH-2000 Neuch\^atel, Switzerland.}
\endauthors

\vskip .3in
\vskip .25in

We show how to systematically derive the rules for bosonization
in two dimensions as a particular case of a duality transformation.
The duality process amounts to gauging the global symmetry of
the original (fermionic) theory, and constraining the corresponding
field
strength $F_{\mu\nu}$ to vanish. Integration over the Lagrange
multiplier,
$\Lambda$, for this constraint then reproduces
the original theory, and integration over the
gauge fields generates the dual theory with $\Lambda$ as the new
 (bosonized)
variable. We work through the bosonization of the Dirac fermion,
the massive and massless Thirring models, and a fermion on a
cylindrical
spacetime as illustrative examples.


\vfill\eject

\section{Introduction}

Quantum field theories in two spacetime dimensions enjoy many special
properties that are not shared by theories in other numbers of dimensions.
These properties have in the past proven to be a mixed blessing.
Although they
have been used to good effect to construct physically-interesting,
 yet solvable,
models, the very dimension-dependence of  these features raises doubts
as to
what lessons can be drawn for higher-dimensional theories.
Their exploration
has undergone a renaissance since the advent of string theory,
 however,
in which two-dimensional field theories play a special role.

One of the two-dimensional tricks which has been most fruitfully
employed
to explore these theories is bosonization. A model involving
two-dimensional
fermions is bosonized when it is re-expressed in terms of an
 equivalent
theory built purely from bosons. Precise rules for various versions
 of this
equivalence have been known for some time now
\ref\linearbos{A. Luther and I. Peschel, \prb{9}{74}{2911}.}
\ref\col{S. Coleman, \prd{11}{75}{2088};\bk
S. Mandelstam, \prd{11}{75}{3026}}
\ref\nonlinearbos{E. Witten, \cmp{92}{84}{455}.}
\linearbos \col \nonlinearbos. The statement of these rules generally
suffers from
the drawback of not being {\it constructive} in character, being instead
couched in terms of the equivalence of a given pair of bose and fermi
theories.\foot\thelit{There are some exceptions to this statement,
for the case
of abelian bosonization. We return to the relation between our
approach and
earlier work in the final section below.} This is particularly true for
nonabelian bosonization, where a simple, systematic, constructive and
unified
presentation of the bosonization rules is still missing
\ref\bosereviews{For a recent review see Y. Frishman and
 J. Sonnenschein, \prep{223}{93}{309}.}
\bosereviews. Our purpose in this note is to provide such a derivation,
for the abelian case.

We do so by showing how to directly construct the equivalent bosonic
theory
using the technique of dualization. Dualization is a particularly
convenient
type of change of variables
\ref\earlydual{See for instance, N.J. Hitchin, A. Karlhede,
U. Lindstrom and M. Rocek, \cmp{108}{87}{535} and
references therein.}
\earlydual, which has recently been found to be useful in relating
otherwise
apparently inequivalent string vacua
\ref\stringdual{T. Buscher, \plb{194}{87}{59}, \plb{201}{88}{466}.}
\stringdual. We show here how the straightforward application of these
transformations to a fermion system provides a `royal road'  to the
bosonized version of the model.

We start with a brief sketch of the dualization technique.
We then illustrate
the ease with which duality transformations deal with bosonization by
considering a series of successively more complicated fermionic models.
After
starting with the simplest case of the bosonization of a free Dirac fermion,
we
turn to the massless and massive Thirring models. Although the initial
discussion assumes a topologically trivial two-dimensional spacetime,
 we then
show how to handle spacetimes with nontrivial homology.

\section{Abelian Bosonization}

Many things in physics go by the name `duality', with the common theme
 being
the equivalence of two superficially quite different formulations of a
particular field theory. Most of the various extant versions of
duality can be thought of as special cases of what we refer to as
duality in
this paper. This duality is a trick for constructing the equivalent
theory
by embedding the model of interest
within a larger gauge theory, which reduces to the original version
 once
the
gauge potential, $A_\mu$, is set to zero. This larger gauge theory is
typically
constructed simply by gauging a global symmetry of the original
 model
\ref\rocekv{M. Rocek and E. Verlinde, \npb{373}{92}{630}.}
\ref\doq{X. de la Ossa and F. Quevedo, \npb{403}{93}{377}.}
\rocekv\ \doq.
The original model is then written as the gauged theory subject
to some
constraint that removes the gauge potential, and this is usually
done
by introducing a dummy variable -- a Lagrange multiplier field,
$\Lambda$ --
whose functional integral requires the gauge field strength to vanish,
$F_{\mu\nu} =0$. (Additional conditions are necessary in topologically
nontrivial spacetimes
\ref\alvarez{E. Alvarez, L. Alvarez-Gaum\'e, J. Barb\'on and
Y. Lozano,  preprint CERN-TH.6991/93.}
\rocekv, \alvarez.)  If $\Lambda$, and then $A_\mu$, are integrated
out --
in this order -- of the gauged version of the theory, then the original
model is
retrieved. The dual version of the theory, on the other hand,
is obtained
by performing the various functional integrals in a different order,
in
particular by integrating out the original field, as well as $A_\mu$.
The
Lagrange multiplier, $\Lambda$, then plays the role of the new, dual,
field. All
of this is best understood in terms of a concrete example, such as we
 give below.

This method has been widely applied to the two-dimensional sigma-models
 which
describe string theory backgrounds \stringdual. Provided that a string
vacuum admits an appropriate symmetry, duality can be used to relate it
to
other, equivalent string solutions. Among the vacua which can be related
in
this way are the toroidal compactifications whose radii are related to
one
another by the duality transformation $R \leftrightarrow \ell_\ssp^2/R$,
with
$\ell_\ssp$ being a string length scale which is of order the Planck size.
It
has also been used to relate black--hole and cosmological string solutions.

\subsection{The Free Dirac Fermion}

Here we apply the duality transformation to the bosonization of a single
 Dirac
fermion which lives in a flat two-dimensional spacetime having the
topology of
\IR${}_2$.\foot\conventions{ Our conventions are: $\ss x^0 = t$,
$\ss x^1 = x$,
$\ss x^\pm = {1\over\sqrt{2}}(x \pm t)$,
 $\ss \eta^{11}= -\eta^{00}= \veps^{01}
=1$, $\ss \gamma_0 = i\sigma_1$, $\ss \gamma_1 = \sigma_2$,
 $\ss \gamma_3
\equiv \gamma^0 \gamma^1 = \sigma_3$, and $\ss \gamma_{\lft} = \hf (1 +
\gamma_3)$.}  For the present purposes we suppose our fields to
fall to zero at
spatial infinity sufficiently quickly to permit the neglect of
 all surface
terms.

We start with the following generating functional for such a particle:
\label\freegenfunction
\eq
Z_\ssf[ J] \equiv  \int \Scd \psi \; \exp\left[ i \int d^2x \;
\left( - \psibar
\dsl \psi + \sum_i  J_i \; \Sco_i(\psi) \right) \right] ,
\eeq
where $J_i=J_i(x)$ represent a set of external fields. These are
 coupled to some
collection of operators, $\Sco_i(\psi)$, which we assume to be invariant
 under
the {\it local} symmetry transformation: $\psi \to e^{i\theta(x)} \, \psi$.
 For
the time being, when we wish to be more concrete, we have in mind the
following
operators:
\label\operators
\eq
\sum_i J_i \; \Sco_i(\psi) = a_\mu \; i \psibar \gamma^\mu \psi +
b_\mu  \; i \psibar \gamma^\mu \gamma_3 \psi,
\eeq
although we generalize to a wider class shortly.

Following the dualization prescription, we gauge the anomaly-free
 vector-like
symmetry, $\psi \to e^{i\theta(x)} \, \psi$, but constrain the
corresponding
field strength to vanish:
\label\gaugedversion
\eq
Z_\ssg[ J] \equiv  \int \Scd \psi \; \Scd A_\mu \; \Scd \Lambda \;
\exp\Biggl[ i
\int d^2x \;  \Biggl( \Scl_\ssf(\psi,J) + i  \psibar \gamma^\mu \psi \;
 A_\mu +
\hf \, \Lambda  \veps^{\mu\nu} F_{\mu\nu} \Biggr) \Biggr]
\Delta[\partial\cdot
A].
\eeq
Here $\Scl_\ssf(\psi,J)$ denotes the lagrangian that appears in
eq.~\freegenfunction, and $\Delta[\partial \cdot A] = \prod_{xt}
\delta[ \partial
\cdot A(x,t)]$ is a functional delta function which imposes the Lorentz
 gauge
condition. Here the functional integration over the Lagrange multiplier,
$\Lambda$, enforces the constraint $F_{\mu\nu} =0$, and in the present
instance
of trivial spacetime topology this, together with the gauge condition,
 allows us
to choose everywhere $A_\mu = 0$. In this way we see the equivalence
 -- up to a
physically irrelevant, $J_i$-independent, normalization constant -- of
expression \gaugedversion\ with our starting point, eq.~\freegenfunction:
 \ie\
$Z_\ssf = Z_\ssg \equiv Z$.

The dual version of the theory is found by instead performing the
functional
integrations over $A_\mu$ and $\psi$, while leaving the integral over
 $\Lambda$
for last. This leads to the following expression for the generating
 functional:
\eq\label\unlabelled
Z[J] = \int \Scd\Lambda \; \exp \Bigl( iS_\ssb[\Lambda,J] \Bigr)
\eeq
where the dual --- or bosonized --- action is defined by:
\label\dualresult
\eq
\exp \Bigl( iS_\ssb[\Lambda,J] \Bigr) \equiv
\int \Scd \psi \; \Scd A_\mu  \;
\exp\Biggl[ i \int d^2x \;  \Biggl( \Scl_\ssf(\psi,J) +
i  \psibar \gamma^\mu
\psi \; A_\mu + \hf \, \Lambda  \veps^{\mu\nu}
F_{\mu\nu} \Biggr) \Biggr]
\Delta[ \partial\cdot A].
\eeq
It only remains to evaluate these integrals.

At this point there are two ways to proceed, depending on which of the
remaining functional integrations are performed first. Since both
choices are
instructive for the more complicated examples which follow, we
describe each in
turn in the present, simpler, context.

The most straightforward way to proceed is to directly evaluate
the integrals
by brute force. This is not so difficult to do for the simple
system we are
considering here. In particular, the result of performing the
fermionic integral
in eq.~\dualresult\ has been known for some thirty years
\ref\schwinger{J. Schwinger, \pr{128}{62}{2425}.}
\ref\jackiw{A useful review with references to the earlier literature is
R. Jackiw, in {\it Relativity, Groups and Topology II},
B. DeWitt and R. Stora, eds. North Holland (1983).}
\schwinger \jackiw. If we renormalize in a gauge invariant way then,
 neglecting
(as always) an irrelevant multiplicative constant, we have:
\eq\label\fermionintegral
\int \Scd\psi \; \exp\left[ i \int d^2x \Bigl( - \psibar \dsl \psi +
i \psibar
\gamma^\mu \psi \; A_\mu \Bigr) \right] = \exp\left[
{i \over 4\pi} \int d^2x \;
F^{\mu\nu} \; \Square^{-1} \; F_{\mu\nu} \right].
\eeq
This expression may be directly applied to eq.~\dualresult, with external
fields coupled to the operators of eq.~\operators, by making the
substitution
$A_\mu \to \hat{A}_\mu \equiv A_\mu + a_\mu + \veps_{\mu\nu} b^\nu$.
 We are
left with the single functional integral:
\eq\label\unlabelled
\exp \Bigl( iS_\ssb[\Lambda,J] \Bigr) = \int \Scd A_\mu  \;
\exp\Biggl[ i \int d^2x \;  \Biggl( {1 \over 4\pi} \hat{F}^{\mu\nu} \;
\Square^{-1} \; \hat{F}_{\mu\nu} + \hf \, \Lambda \veps^{\mu\nu} F_{\mu\nu}
\Biggr) \Biggr] \Delta[ \partial\cdot A].
\eeq

Next comes the integration over $A_\mu$. To this end, we first perform
the
following change of variables: $A_\mu \to \varphi$, with $A_\mu \equiv
\veps_{\mu\nu} \; \partial^\nu \varphi$. The Jacobian for this
transformation is
an irrelevant constant, so we have $\Scd A_\mu \; \Delta[\partial \cdot A] =
\Scd \varphi$. The resulting $\varphi$ integration is Gaussian,
with a saddle
point at $\varphi = \varphi_c$, with:
\eq\label\unlabelled
\Square \, \varphi_c + \pi \, \Square \, \Lambda + \partial^\mu (b_\mu +
\veps_{\mu\nu} \, a^\nu) = 0.
\eeq
Using this to evaluate the integral over $\varphi$, gives the standard
bosonic
result:
\label\theanswer
\eq
\Scl_\ssb(\Lambda,a,b) = -{ \pi \over 2} \; \partial_\mu \Lambda \,
\partial^\mu
\Lambda + \partial_\mu \Lambda \, b^\mu + \veps^{\mu\nu} \; \partial_\mu
\Lambda
\, a_\nu + B ,
\eeq
where $B$ is an arbitrary $\Lambda$- and $b_\mu$-independent constant.
The precise value that is taken by the constant $B$ depends on the
details of
how the path integral is regularized and renormalized. It is
conventional to
choose it to ensure that the vacuum energy vanishes, in which case
$B = 0$ at
the classical level. It is noteworthy that in obtaining expression
 \theanswer,
the potentially nonlocal terms in Schwinger's result,
 eq.~\fermionintegral,
cancel against those that arise when the integral over the field
 $\varphi$ is
performed.

In terms of the canonically-normalized scalar variable,
 $\phi = \sqrt{\pi} \,
\Lambda$, the dual lagrangian takes its standard form:
\label\canresult
\eq
\Scl_\ssb(\phi,a,b) = - \hf \; \partial_\mu \phi \, \partial^\mu \phi
+ \nth{\sqrt{\pi}} \; \partial^\mu \phi \, b_\mu + \nth{\sqrt{\pi}} \;
\veps^{\mu\nu} \,\partial_\mu \phi \, a_\nu .
\eeq

Comparing the coefficients of $a_\mu$ and $b_\mu$ in eqs.~\operators\
and
\canresult\ shows that the currents in these two theories are related
 by
\eq\label\unlabelled
i \psibar \gamma^\mu \psi \leftrightarrow - \, \nth{\sqrt{\pi}}
\;\veps^{\mu\nu}
\, \partial_\nu \phi, \qquad \hbox{and:} \qquad
i \psibar \gamma^\mu \gamma_3 \psi \leftrightarrow \nth{\sqrt{\pi}} \;
\partial^\mu \phi .
\eeq
This ultimately justifies the equivalence of the duality transformation
as
described above with bosonization as it is conventionally formulated.

\vfill\eject
\subsection{The Dirac Fermion Revisited: Symmetry Arguments}

The previous derivation of eq.~\theanswer\ from eq.~\dualresult\ has
the
advantage of being conceptually straightforward. Yet it also has the
disadvantage of relying on being able to explicitly perform all of the
relevant
functional integrals. This makes it less easy to apply to more
complicated
fermionic models. As a result we here rederive eq.~\theanswer\ from
eq.~\dualresult\ in a way that is more amenable to generalization.

This alternative approach starts by first performing the integral over
the gauge
potential, $A_\mu$. To proceed we therefore write the gauge condition
as a
functional Fourier transform:
\eq\label\unlabelled
\Delta[\partial
\cdot A] = \int \Scd \omega \; \exp\left[ i \int d^2x \; \omega \;
\partial\cdot A
\right].
\eeq
The integral over $A_\mu$ is now unconstrained, and since the action is
linear
in this variable, we are led to the result:
\label\newdualresult
\eq
\exp \Bigl( iS_\ssb[\Lambda,J] \Bigr) = \int \Scd \psi \;\Scd \omega \;
\exp\Bigl( i S_\ssf[\psi,J] \Bigr) \; \Delta\Bigl[ i \psibar\gamma^\mu \psi +
\veps^{\mu\nu} \; \partial_\nu \Lambda - \partial^\mu \omega \Bigr].
\eeq

The trick is to now indirectly determine the result of the remaining
integrations. We wish to do so by taking advantage of the symmetries of
the
problem --- in particular the axial $U_\ssa(1)$ symmetry. Consider, then,
as a
first case, coupling an external field only to the axial-current operator:
$J_i \Sco_i = i\psibar \gamma^\mu \gamma_3 \psi \; b_\mu$. Classically,
the
lagrangian of eq.~\gaugedversion\ then has the local axial $U_\ssa(1)$
symmetry,
$\psi \to e^{i\alpha \gamma_3} \psi$ and $b_\mu \to b_\mu + \partial_\mu
\alpha$, in addition to the vector symmetry which we previously gauged
using the
gauge potential $A_\mu$. This axial symmetry does not survive
 quantization,
however   \jackiw, as may be seen explicitly, \eg, by using a
 point-splitting
regularization.  In this scheme the regularized axial and vector
currents have
the following transformation properties:
\eq\label\unlabelled
i \psibar \gamma^\mu \gamma_3 \psi \to i \psibar \gamma^\mu \gamma_3 \psi
  +
{1\over \pi} \; \partial^\mu \alpha, \qquad \hbox{and:} \qquad  i
\psibar
\gamma^\mu \psi \to i \psibar \gamma^\mu \psi  - {1\over \pi} \;
 \veps^{\mu\nu}
\; \partial_\nu \alpha .
\eeq
As a result, the regularized lagrangian $\Scl_\ssf(\psi,A,b) \equiv -
\psibar \dsl \psi +i \psibar \gamma^\mu \psi \; A_\mu + i \psibar
 \gamma^\mu
\gamma_3 \psi \; b_\mu$ transforms according to
\label\variation
\eq
\Scl_\ssf \to \Scl_\ssf + {1\over \pi} \; \partial_\mu \alpha \,
 (b^\mu +
\veps^{\mu\nu} A_\nu).
\eeq

Now, although the $b_\mu$-dependent term in this equation can be
cancelled by a
local counterterm involving just the fields $\psi, A_\mu$ and $b_\mu$ ---
\ie\ by: $\Scl_{\rm ct} = - {1 \over 2 \pi} \; b_\mu b^\mu$ --- the
second term
cannot. It  represents the familiar $U_\ssa(1)$ axial anomaly.
The key point is
that this anomaly {\it can} be cancelled by a term in the gauged
action of eq.~\gaugedversion, however, because of the inclusion there
 of the
extra,  Lagrange multiplier, field $\Lambda$. It is cancelled if this
field
acquires the  transformation property $\Lambda \to \Lambda +
 \alpha /
\pi$. In this regard the Lagrange-multiplier term in the lagrangian,
 $\hf \,
\Lambda \veps^{\mu\nu} F_{\mu\nu}$, plays the role of a Green-Schwarz,
or
Wess-Zumino, term for cancelling the fermion axial anomaly.

The upshot of all of this is that the bosonic lagrangian,
$\Scl_\ssb(
\Lambda,b)$, as defined by eq.~\dualresult, transforms under the
 complete axial
transformations, $\Lambda \to \Lambda +
\alpha /\pi$ and $b_\mu \to b_\mu
+ \partial_\mu \alpha$, in such a way as to ensure that the
quantity $\Scl_\ssb(
\Lambda,b) - {1\over 2\pi} \; b_\mu b^\mu$ is invariant.
It follows that this
quantity can depend on $\Lambda$ and $b_\mu$ only through the
covariant
derivative $D_\mu \Lambda \equiv \partial_\mu \Lambda - \nth{\pi}
\; b_\mu$.

The form of this invariant term of the lagrangian can now be pinned
down by
using the following argument. Differentiating eq.~\newdualresult\ with
respect to $b_\mu$ gives:
\label\dSBdb
\eq \eqalign{
{\delta S_\ssb \over \delta b_\mu} &= \avg{ i \psibar \gamma^\mu
\gamma_3 \psi
} \cr
&= \avg{ i \psibar \gamma_\nu \psi} \; \veps^{\nu\mu} \cr
&= \partial^\mu \Lambda -\veps^{\mu\nu} \; \avg{\partial_\nu \omega}
.\cr}
\eeq
Here $\avg{X}$ denotes the average of $X$ over the variables $\psi,
 A_\mu$ and
$\omega$:
\eq\label\unlabelled
\avg{X} \equiv {\int \Scd\psi \Scd A \Scd\omega \; X \; e^{iS_\ssg}
\over \int \Scd\psi \Scd A \Scd\omega \; e^{iS_\ssg}}.
\eeq

We finally require an expression for $\avg{\omega}$ in terms of
the variables $\Lambda$ and $b_\mu$. As we argue below, this quantity
vanishes.
The transformation property of $S_\ssb$ under axial transformations,
 together
with the functional derivative of eq.~\dSBdb, then implies
\eq\label\unlabelled
\Scl_\ssb(\Lambda,b) = -{ \pi \over 2} \; \left(\partial_\mu \Lambda -
\nth{\pi}
\; b_\mu \right) \, \left(\partial^\mu \Lambda - \nth{\pi} \;
 b^\mu \right)
+ {1\over 2\pi} \; b_\mu b^\mu + B.
\eeq
This reproduces the result of eq.~\theanswer\ for the special case
 where $a_\mu
= 0$. The coupling to $a_\mu$ is also easily included.
The simplest procedure is
to remark that in the original fermion theory of eq.~\freegenfunction,
 the
identity $\gamma^\mu \gamma_3 = \gamma_\nu \veps^{\nu\mu}$ implies
that the
fields $a_\mu$ and $b_\mu$ only appear \via\ the combination
 $b_\mu +
\veps_{\mu\nu}\, a^\nu$. We must therefore add the term
 $\veps^{\mu\nu} \;
\partial_\mu \Lambda \, a_\nu$ to the above result. (Notice that,
 keeping in
mind the transformation rule, $\Lambda \to \Lambda + \alpha/\pi$,
this term is
precisely what is required to reproduce the axial anomaly of the
 original
fermion theory.) We reobtain in this way the desired expression,
eq.~\theanswer.

The missing step in the above is the establishment of the result
 $\avg{\omega} =
0$. This can be argued from the CP-transformation properties of the
functional
integrand. The action $S_\ssg$ is invariant under the following
 substitutions:
$\omega \to - \omega$, $\psi \to \sigma_2 \psi^*$, $A^\mu \to -
{P^\mu}_\nu
A^\nu$ and $b^\mu \to + {P^\mu}_\nu b^\nu$, where the
parity-transformation
matrix is ${P^\mu}_\nu = \diag{+1,-1}$. Since $\omega$ is odd
under this
transformation, its average value must be zero.
This is because the contribution
of any particular field configuration to $\avg{\omega}$ is
 systematically
cancelled by the contribution of its CP-conjugate configuration.

An interesting consequence of the bosonic tranformation rule,
 $\Lambda \to
\Lambda + \alpha/\pi$, is that $\Lambda$ is itself a periodic variable.
That is,
since an axial rotation through a complete turn of $2\pi$ radians leaves
the
fermion field unchanged, we must identify $\Lambda \cong \Lambda + 2$.
A
stronger periodicity condition, $\Lambda \cong \Lambda +1$, becomes
 possible if
we only demand equivalence up to a lorentz transformation, since an
axial
rotation through $\pi$ radians takes $\psi$ to $- \psi$, which equals
a spatial rotation through 360${}^o$.

We next turn to more complicated applications.

\subsection{Masses and the Thirring Model}

It is straightforward to extend the previous construction to the massless
and
massive Thirring models, for which the fermion action, $S_\ssf$, is
supplemented
by additional terms, $\Delta S_g$ and $\Delta S_m$, which respectively
 contain
a four-fermion contact interaction, and a coupling of the
chirality-breaking
operators, $\psibar \gamma_{\lft,\rht} \psi$, to external currents.
That is:
\label\newterms
\eq
\Delta \Scl_g = - {g^2 \over 2} \, (\psibar \gamma^\mu \psi)
\; (\psibar \gamma_\mu \psi)  \qquad \hbox{and} \qquad
\Delta \Scl_m = - m(x) \; \psibar \Pl \psi - m^*(x) \; \psibar \Pr \psi .
\eeq
The special case of a spatially constant $m$ simply corresponds to massive
fermions.

The four-fermi term, $\Delta S_g$, can be handled as a special case of the
treatment of the coupling to the external field $a_\mu$. This can be seen
by using the trick of `uncompleting the square'
\ref\uncompleting{Y. Nambu and G. Jona-Laisinio, \pr{122}{61}{231};
\bk
D. Gross and A. Neveu, \prd{10}{74}{3235}.}
\uncompleting, in which $-\frac{g^2}{2} \, (\psibar \gamma^\mu \psi)
(\psibar
\gamma_\mu \psi)$ is rewritten as $- \hf \, c_\mu c^\mu + i g \, c_\mu \,
\psibar \gamma^\mu \psi$. The original four-fermion form is then
recovered by
performing the Gaussian integration over the dummy field $c_\mu$.
Performing the
integrals over $\psi$ and $A_\mu$ as in the previous section,
followed by the
integral over $c_\mu$, then gives the following additional
contribution to the
bosonic action:
\label\thirringterm
\eq - {g^2 \over 2} \; \partial_\mu \Lambda \partial^\mu \Lambda. \eeq
Its net effect is therefore to simply change the relation between
$\Lambda$ and
the canonically-normalized variable, $\phi$, which is now given by
$\phi =
\sqrt{\pi + g^2} \; \Lambda$. (It is clear from this expression that the
stability of the
bosonic theory requires  $g^2 > -\pi$.) This shows up in a change in
the
normalization of the expressions for the currents, which become:
$i\psibar
\gamma^\mu \psi \leftrightarrow  - \, \nth{\sqrt{\pi + g^2}} \;
\veps^{\mu\nu} \,
\partial_\nu \phi$ \etc.

Since it changes the relation between $\phi$ and $\Lambda$,
this $g$-dependent
change of normalization has a geometrical interpretation in
terms of the
periodicity  of the bosonic variable, $\Lambda$. In string theory it
is
conventional to define the radius, $R$, of the space parameterized
 by $\Lambda$
by writing the kinetic term as $- {R^2 \over 4\pi \alpha'} \;
\partial_\mu
\Theta \partial^\mu \Theta$, where the variable $\Theta$ is defined to
 have the
periodicity $\Theta \cong \Theta + 2 \pi$. Given the basic equivalence
$\Lambda \cong \Lambda + 2$, we see that $\Theta \equiv \pi \Lambda$.
($\alpha'$ is the square of the fundamental string length scale, which
is of
order the Planck length.) With this choice we see that the radius so
defined is
related to the Thirring model coupling according to $R^2 = 2\alpha' \;
\left(1 +
{g^2 \over \pi} \right) $.

The $m(x)$-dependent terms can also be included along the same lines as
for the
fermion kinetic terms. In this case the axial-$U_\ssa(1)$ transformation
properties of bosonic lagrangian are:
\label\massvariation
\eq
{\delta \over \delta \alpha} \left[ \Scl_\ssb - \nth{2\pi} \; b_\mu
b^\mu \right] = - 2 i \, m(x) \; \avg{ \psibar \Pl \psi}  + 2
i\, m^*(x) \; \avg{ \psibar \Pr \psi}.
\eeq
Integration of this expression to obtain $S_\ssb$ requires an expression
for
the expectation value $F[\Lambda] \equiv \avg{\psibar \Pl \psi}$, and its
complex conjugate $F^*[\Lambda] = \avg{\psibar \Pr \psi}$ as functionals
 of the
bosonic fields $\Lambda$, $a_\mu, b_\mu$, and $m$. The $\Lambda$-dependence
 can
be obtained by using its transformation properties under the chiral
symmetry: $F
\left[\Lambda + \nth\pi \,\alpha \right] \equiv e^{2i\alpha} \; F[\Lambda]$,
for
all $\Lambda$ and $\alpha$. The most general solution to this functional
identity is: $F[\Lambda] = A \exp(2\pi i \, \Lambda)$, where $A$ is
invariant under local chiral transformations.
(Similarly $\avg{\psibar \Pr \psi} =
 A^* \exp(-2\pi i \,
\Lambda)$.)
 In principle chiral invariance  permits $A$ to depend on the variables
$\Lambda$ and $b_\mu$, provided that they only appear through invariant
combinations such as $(\partial_\mu \Lambda - {1 \over \pi} \; b_\mu)^2$.
However this possibility can be ruled out by considering the functional
derivative $\delta S_\ssb/\delta b_\mu$, for
which the arguments of the previous section again imply $\delta
S_\ssb/\delta b_\mu = \partial^\mu \Lambda$. As a result we find
that $A$ must be  $\Lambda$- and $b_\mu$-independent.

Using this in eq.~\massvariation, allows us to infer the $m$-dependent
chiral-symmety breaking terms which appear in $S_\ssb$.
Switching once more to
canonically-normalized variables, we find the general bosonized
form:
\label\thewholeschmeer
\eq
\Scl_\ssb(\phi,a,b,m) = - \hf \, \partial_\mu \phi \, \partial^\mu \phi +
{\beta \over 2\pi} \; \partial_\mu \phi \; (b^\mu + \veps^{\mu\nu} a_\nu)
- A \left[ m \exp (i\beta\phi) + \hc \right] + B.
\eeq
The parameter $\beta$ appearing in this formula is related to the
Thirring-model coupling by $\beta = 2 \pi/\sqrt{\pi + g^2}$. Again both
 $A$ and
$B$ are constants which depend on how the path integral is renormalized.
The
conventional choice of vanishing vacuum energy leads to a relation
between $B$ and $A$ at the classical level -- \eg\ for $m=m^*$, $B = -2mA$.
 The
value taken for $A$ depends on the renormalization condition that is
used to
define the composite operators $\exp(i \beta \phi)$ and
$\psibar \Pl \psi$.

As for the simpler case treated earlier,
 eq.~\thewholeschmeer\ reproduces the
results of earlier workers \col.

\section{Bosonization on the Cylinder}

As our next application we apply our procedure to bosonization on
 spacetimes
with nontrivial topology. For the purposes of illustration we consider
 here the
simplest possible example: a single free Dirac fermion on a cylinder.
 We see
no obstacle to the application of our method also to
Euclidean spaces of more complicated topology,
such as to the Riemann surfaces
which appear in string theories
\ref\stringtopexamples{L. Alvarez-Gaum\'e, G. Moore and C. Vafa.
\cmp{106}{86}{1}.}
\ref\oscar{P. Griffin and O. Hern\'andez, \npb{356}{91}{287};
\ijmp{7}{92}{1233}.}
\stringtopexamples. Some of the issues we encounter in this section
have also arisen in applications of duality to string theory, such as in
ref.~\rocekv\ \alvarez\ \oscar, as well as in earlier work with the
Thirring
and Schwinger models
\ref\smoncylinder{D. Wolf and J. Zittartz, {\it Zeit. Phys.}
{\bf B51} (1983) 65;\bk N. Manton \anp{159}{85}{220};
\bk J.E. Hetrick and Y. Hosotani,
\prd{38}{88}{2621}.}
\smoncylinder.

Consider, therefore, a cylindrical spacetime for which the spatial
coordinate
is periodic with period $L$: $x \cong x + L$ (for all $t$).
The new feature in
this example is the existence of the nontrivial homology cycle, $\ell$,
consisting of curves which wind around the cylinder.
 A two-dimensional
Dirac fermion, $\psi$, can have several choices of boundary
 conditions,
or spin structures, according to whether $\psi_\lft$ and $\psi_\rht$
are
periodic or antiperiodic around such curves:
\label\boundaryconditions
\eq
\psi(x+L,t) = \Bigl[ (-)^{n_\lft} \Pl + (-)^{n_\rht} \Pr \Bigr] \;
 \psi(x,t).
\eeq
Here the spin structures are labelled by the pair of integers,
$(n_\lft,n_\rht)$, modulo 2. In what follows we restrict ourselves, for
simplicity, to nonchiral boundary conditions for which $n_\lft = n_\rht
\equiv
n$. We wish to determine the bosonic theory which is dual to a free
 Dirac
fermion, $\psi_n$, which satisfies these boundary conditions.

We therefore consider again the fermionic generating function:
\label\newfreegenfn
\eq
Z_{n,\ssf}[a,b] \equiv \int \Scd\psi_n \; \exp\left[ i \int d^2x \;
\left( -
\psibar_n \dsl \psi_n +  i \psibar_n \gamma^\mu \psi_n \; a_\mu +
 i \psibar_n
\gamma^\mu \gamma_3 \psi_n \; b_\mu \right) \right] ,
\eeq

As before, to dualize we must regard the above expression as coming
from a
gauge theory that is subject to a gauge-invariant constraint.
A key difference
from the previous section, however, is that this time it is {\it not}
sufficient
to use the constraint $F_{\mu\nu}=0$ to achieve this end, since this
constraint,
taken together with a homogeneous gauge (like Lorentz gauge), can only
 ensure
that the spatial component, $A_1$, of the gauge potential is a
spacetime-independent  constant. This constant cannot be gauged away
using a
gauge transformation $A_\mu \to A_\mu + \partial_\mu \theta$,
for which the
gauge group element, $g(x) = e^{i \theta(x)}$, is a periodic
function of $x$. In
fact, the $x$-independent part of $A_1$ can only be shifted by `large'
 gauge
transformations, which satisfy $\theta(x+L,t) = \theta(x,t) + 2 \pi k$,
 with $k$
an integer, so $A_1$ can be made to take values on a circle of
 circumference
$2\pi/L$. A convenient way to characterize the constant mode of
 $A_1$ is through
the Wilson loop, given by $W[A] \equiv \int_\ell A_\mu \; dx^\mu =
 \int_0^L A_1
\; dx$. Clearly  if both $F_{\mu\nu} = 0$ {\it and} $W[A] = 0$ vanish,
then
$A_\mu = 0$ up to a periodic gauge transformation.

We therefore adopt the following gauged generating functional:
\label\newgaugedversion
\eq \eqalign{
Z_{n,\ssg}[a,b] &\equiv  \int \Scd \psi_n \; \Scd A_\mu \; \Scd \twL
\exp\Biggl[ i \int d^2x \;  \Biggl( \Scl_\ssf(\psi_n,a +A,b) + \hf \,
 \twL
\veps^{\mu\nu}  F_{\mu\nu}\Biggr) \Biggr] \cr
& \qquad \qquad \qquad \qquad \qquad \qquad \qquad \qquad \qquad
\times \;
\Delta[\partial\cdot A] \; \Delta_t(W[A]). \cr}
\eeq
Several features of this expression bear explicit mention:
\topic{1}
The main new feature in eq.~\newgaugedversion\ is the additional
delta
function
$\Delta_t (W[A])$. Here the subscript `$t$' indicates that this
functional delta
function is to be imposed separately at each instant of time only,
since $W([A])
= \int_0^L dx \; A_1$ is itself independent of $x$. It will prove
to be
convenient to write this constraint as a functional Fourier
 transform:
\label\unlabelled
\eq
\Delta_t (W[A]) = \int \Scd u_t \; \exp \left[ i \int d^2x \; u_t \;
 A_\mu
s^\mu \right], \eeq
where $s^\mu = \delta^\mu_1$ is a constant unit vector pointing in the
 $x$
direction,  and the subscript `$t$' on the Lagrange-multiplier field,
$u_t$,
 is
meant to emphasize that it is a function of $t$ only: $u_t = u(t)$.

This new contribution serves two separate purposes. Firstly,
it provides
the
additional condition, as discussed above, that is required in
 order to constrain
the gauge potential to vanish, and so to ensure the equivalence of
eq.~\newgaugedversion\ with our starting point, eq.~\newfreegenfn:
 $Z_{n,\ssf} =
Z_{n,\ssg} \equiv Z_n$. It also serves to remove the freedom to perform
`large'
gauge transformations, which would otherwise be unfixed since this
 freedom is
not constrained by the imposition of Lorentz gauge, $\partial
 \cdot A = 0$.

\topic{2}
Notice that both of the bosonic integration variables in
 eq.~\newgaugedversion\
must be strictly periodic in $x$. This is true for both components
of $A_\mu$
since these must have  the same behaviour as does the derivative
$\partial_\mu$.
Similarly, the Lagrange multiplier field, $\twL$, (the tilde is
included here
for later notational  convenience) must also be periodic since
 otherwise its
interaction lagrangian, $\hf \, \twL  \veps^{\mu\nu} F_{\mu\nu}$,
itself would
not be  single-valued on the cylinder.
\endtopic

As in the previous section, the dual version of the theory is
 found by
performing all of the functional integrals with the exception
of that over
$\twL$. This gives $Z_n[a,b] = \int \Scd\twL \; \exp \Bigl(
iS_\ssb[\twL, a,b] \Bigr)$, with
\label\newbosedef
\eq \eqalign{
\exp \Bigl( iS_\ssb[\twL,a,b] \Bigr) &\equiv \int
\Scd \psi_n \; \Scd A_\mu \;
\Scd u_t \; \exp \Biggl[ i \int d^2x \;
 \Biggl( \Scl_\ssf(\psi,A+a,b)
 \cr
& \qquad\qquad  + \twL \veps^{\mu\nu} \partial_\mu A_\nu +
 u_t A_\mu s^\mu
\Biggr) \Biggr] \; \Delta[\partial \cdot A]. \cr}
\eeq

The integration over $u_t$ is quite easy to deal with.
This is because, apart
from a surface term, the quantities $u_t$ and $\twL$
only enter into the above expression through the combination
$\veps^{\mu\nu}
\partial_\nu \twL + u_t s^\mu$. As a result the $x$-independent
mode of $\twL$
and $u_t$ are redundant variables, and $u_t$ may be completely
removed
from the integrand of eq.~\newbosedef\ by redefining
\label\redefinition
\eq
\Lambda \equiv \twL - f(t),
\eeq
with ${df\over dt} = u_t$. This defines $\Lambda$ up to the
ambiguity of an
additive constant. With this choice the remaining integration
over $u_t$ just
contributes a field-independent overall factor to $Z_n$,
 which can be
ignored. The ability to completely remove $u_t$ in this way shows
that it is
actually sufficient to impose $W[A]=0$ at any one time, $t$, rather
than
independently for each $t$, as was done in the above.

We now must grapple with the evaluation of the various functional
integrations.
We do so below, treating in turn the cases of a periodic and
 antiperiodic
fermion.

\subsection{Periodic Fermions}

Consider first the case where the fermion is periodic in the spatial
 direction
($n=0$): $\psi(x+L,t) = \psi(x,t)$. In this case we proceed
 \`a la Schwinger,
and explicitly integrate out first the fermion, and then the
gauge potential.
(Notice that even though the fermions are periodic, we need not
worry about
fermion zero modes in the path integral, since the potentially
dangerous mode
$\psi = \hbox{(constant)}$ is not normalizable on an infinitely
long
cylinder.)

We therefore repeat the steps of the previous section, starting
 from the
basic result for the fermion integration, eq.~\fermionintegral,
 which states:
\eq\label\newfermionintegral
\int \Scd\psi_p \; \exp\left[ i \int d^2x \Bigl( - \psibar \dsl
\psi + i
\psibar \gamma^\mu \psi \; A_\mu \Bigr) \right] = \exp\left[
{i \over 4\pi} \int
d^2x \; F^{\mu\nu} \; \Square^{-1}_p \; F_{\mu\nu} \right].
\eeq
The subscripts `$p$' on the fermion measure, and the Green's
function for
$\Square$, indicate that the corresponding quantities are periodic
about the
cylinder's circular direction.

Since the argument proceeds precisely as before, we only sketch
 the main points
again here, with an emphasis on those features which are special
for the
cylindrical geometry. As before, we change variables to $A_\mu =
\veps_{\mu\nu}
\; \partial^\nu \varphi$, where the periodicity of the gauge potential,
$A_\mu$,
ensures the same for the scalar field $\varphi$. Performing the same
integrations
as in the earlier section, we find the same result as in eq.~\theanswer:
\label\boseresult
\eq
\Scl_\ssb(\Lambda,a,b) = -{ \pi \over 2} \; \partial_\mu \Lambda \,
\partial^\mu \Lambda + \partial_\mu \Lambda \, b^\mu + \veps^{\mu\nu} \;
\partial_\mu \Lambda \, a_\nu .
\eeq
Clearly a periodic fermion bosonizes to a periodic boson, with the same
generating function as was obtained when spacetime was \IR${}_2$.

It is crucial in obtaining this result that the fermions really are
 periodic.
This is because in obtaining eq.~\boseresult\ from eq.~\newbosedef,
 there is a
cancellation between two potentially nonlocal terms, having the
form
\label\cancellation
\eq
{1 \over 2\pi} \; \left[ \partial^\mu (a_\mu +
\veps_{\mu\nu} \, b^\nu) \;
\left( \Square^{-1}_p - \Square^{-1}_p \right) \;
\partial^\lambda (a_\lambda +
\veps_{\lambda\rho} \, b^\rho) \right].
\eeq
For this cancellation to work, both of these Green's functions
 must satisfy the
{\it same} boundary conditions, as they do since both are periodic
around the
cylinder. But the boundary conditions of the first Green's
function in
eq.~\cancellation\ are those of the scalar field, $\varphi$ --- which
 is periodic
since $A_\mu$ is --- while the second Green's function in
\cancellation\ has the
boundary conditions of the fermion, $\psi$. The construction must
evidently
be different for antiperiodic fermions, to which we return shortly.

\vfill\eject
\subsection{Winding Sectors}

In the previous section we were led to a dual form which involved
a boson field,
$\Lambda$, which was strictly periodic about the cylinder.
Since $\Lambda$ is
itself periodic (in the target-space sense) under shifts by
an integer,
it can satisfy boundary conditions for
which $\Lambda(x+L,t) = \Lambda(x,t) + k$, for $k$ an integer.
 We pause here to
determine the fermionic system which corresponds to these more
general boundary
conditions.

Consider, therefore, the fermion generating functional in which
all of the
fermion field configurations are restricted to have a definite
integer charge,
$q$. That is, define $Z^q_\ssf$ by inserting the following
functional delta
function into the generating functional of eq.~\newfreegenfn:
\eq \eqalign{
\Delta_t[Q - q] &\equiv \int \Scd v_t \; \exp\left[ i \int dt
\; v_t \, \Bigl(
Q - q \Bigr) \right] \cr
&= \int \Scd v_t \; \exp\left[ i \int d^2x \; v_t \,
\Bigl( i \psibar \gamma^0
\psi  - {q \over L} \Bigr) \right]. \cr}
\eeq
As before the subscript on the function $v_t$ is to emphasize
that it is a
function of $t$ only.

Bosonization of the generating functional in the presence of
this delta function
may be carried out as in the previous section. It leads to  precisely
 the same result
as before, eq.~\boseresult, with the following three modifications:
 ($i$) the
axial field $b_\mu$ must be shifted to $b_\mu - v_t \,
\delta^1_\mu$; ($ii$)
the additional term, $- q \int dt \; v_t $, must be added to
 $S_\ssb$; and
($iii$) the functional integral over $v_t$ must be performed.
The net result for the bosonic action becomes:
\label\qboseresult
\eq \eqalign{
e^{iS^q_\ssb[\Lambda,a,b]} &\equiv \int \Scd v_t \; \exp
\left\{ i
\int d^2x \; \left[\Scl_\ssb(\Lambda,a,b) - v_t \left(
{\partial \Lambda \over
\partial x} + {q \over L} \right) \right] \right\} \cr
&= \exp \left\{ i\int d^2x \; \left[ -{ \pi \over 2} \;
\partial_\mu \Lambda \,
\partial^\mu \Lambda + \partial_\mu \Lambda \, b^\mu +
\veps^{\mu\nu} \;
\partial_\mu \Lambda \, a_\nu  \right] \right\} \cr
&\qquad\qquad\qquad\qquad\qquad\qquad \times \;
\Delta_t[\Lambda(L,t) -
\Lambda(0,t) + q]. \cr}
\eeq

This gives the usual expression, eq.~\boseresult, for the dual
action, but with
the new feature that the integration over $\Lambda$ is only to
 be performed over
the configurations which wind $-q$ times around the cylinder.
We see in this way
the one-to-one correspondence between the sectors of definite
charge in the fermionic
theory, and the sectors with a given winding number in the
bosonic theory.

\vfill\eject
\subsection{Antiperiodic Fermions}

In order to handle antiperiodic fermions we use the artifice of
trading the
new boundary condition for the previous one. That is, we recast
the
fermionic theory by defining a new Dirac spinor $\psi$, by
$\psi(x,t) \equiv
e^{i\alpha_n \gamma_3} \; \psi_n(x,t)$, with $\alpha_n =
 n \pi x/L$. The
boundary conditions of eq.~\boundaryconditions\ for $\psi_n$,
 then imply that
$\psi$ must simply be periodic: $\psi(x+L,t) = \psi(x,t)$.

The information that is encoded in the boundary condition does
 not disappear,
however, since this field redefinition does {\it not} leave
 the lagrangian
invariant. The change in the lagrangian can be most simply
inferred by
recognizing that $\psi$ and $\psi_n$ are related by a local
axial
transformation, and we have determined the transformation
properties of the
bosonic lagrangian under such axial transformations in the
previous sections.
In particular, we know that the effect of performing a
transformation of the
form $\psi \to e^{i\alpha \gamma_3} \; \psi$ in the definition,
\newbosedef, of
$S_\ssb[\Lambda,a,b]$, has the following three effects:
($i$) those which can be
undone by performing the compensating transformations
$b_\mu \to b_\mu
+ \partial_\mu \alpha$ and $\Lambda \to \Lambda +
\alpha/\pi$; ($ii$) those
which can be cancelled by the addition of the counterterm
${1 \over 2 \pi} \;
b_\mu b^\mu$; and ($iii$) those due to the axial anomaly,
which are reproduced
by the term $\veps^{\mu\nu} \; \partial_\mu \Lambda \, a_\nu$.

That is to say, the bosonic result, $\Scl_{n,\ssb}(\Lambda,a,b)$,
for
antiperiodic fermions is related to the corresponding result,
eq.~\boseresult,
for periodic fermions, $\Scl_\ssb(\Lambda,a,b)$, by:
\label\relation
\eq
\Scl_{n,\ssb}(\Lambda,a,b) - {1 \over 2 \pi} \; b_\mu b^\mu -
\veps^{\mu\nu} \;
\partial_\mu \Lambda \, a_\nu = \Scl_\ssb(\Lambda',a,b') -
{1 \over 2 \pi} \;
b'_\mu b'^\mu - \veps^{\mu\nu} \; \partial_\mu \Lambda' \, a_\nu ,
\eeq
in which $\Lambda' \equiv \Lambda - \alpha_n/\pi$ and $b'_\mu
\equiv b_\mu -
\partial_\mu \alpha_n$. This implies:
\label\eitherbc
\eq
\Scl_{n,\ssb} = -{ \pi \over 2} \; \partial_\mu \Lambda' \,
\partial^\mu \Lambda' + \partial^\mu \Lambda' \, b'_\mu +
 \veps^{\mu\nu} \;
\partial_\mu \Lambda' \, a_\nu + {1 \over \pi} \; \partial^\mu
 \alpha_n \Bigl(
b'_\mu + \veps_{\mu\nu} \; a^\nu \Bigr) + \hbox{(constant)}.
\eeq
Notice that the field, $\Lambda'$, which appears in this
expression is no
longer periodic around the cylinder since the periodicity
of $\Lambda$ implies
\label\newbc
\eq
\Lambda'(x+L,t) = \Lambda'(x,t) - n.
\eeq
Thus, we are ultimately led to the boundary condition for
 $\Lambda$ that
would be expected for antiperiodic fermions from the usual
operator
correspondence $\psi \leftrightarrow \exp(i \pi \Lambda)$.

Notice that this boundary condition is also what was encountered
for the
bosonization of periodic fermions subject to the constraint $Q = n$.
 The
functional integrands are not identical, however, due to the
appearance in
the present case of the additional terms:
\eq
{1 \over \pi} \; \partial^\mu \alpha_n \Bigl(
b'_\mu + \veps_{\mu\nu} \; a^\nu \Bigr) =
{n \over L} \Bigl( b'_1 + a^0).
\eeq
Since the electric charge in the fermionic theory is
obtained from the
generating functional by differentiating with respect to the
$x$-independent
mode of $b_1$ or $a^0$, these additional terms are precisely what is
required to
cancel the charge that would otherwise be implied by the boundary
 condition
of eq.~\newbc.

\section{Conclusions}

Our purpose here has been to show how the two-dimensional technique of
bosonization can be considered to be a special case of the wider class
of
relationships amongst two-dimensional theories that has emerged
from the study
of string theories. In particular, we show here how the usual rules
of abelian
bosonization follow systematically as a particular application of a
 duality
transformation. Besides the intrinsic interest of placing the
bosonization
technique within this wider framework, we regard our new perspective
on
bosonization as being useful inasmuch as it permits a constructive
determination of the bosonic counterpart of any given fermionic theory.
 This
should permit a more systematic determination of the bosonization
rules in
more complicated systems, such as for fermions on arbitrary Riemann
surfaces, or
for the nonabelian bosonization of $N$ majorana fermions,
for which the nonabelian duality technique of \doq\ should
 give
the Wess-Zumino-Witten (WZW) model after bosonization.

Other path integral approaches to abelian bosonization
\ref\damgaard{P.H. Damgaard, H.B. Nielsen and R. Sollacher,
\npb{385}{92}{227};
\plb{296}{92}{132}.}
exist \damgaard,  which hinge on the introduction of a
`collective field' through the performance of a local chiral
transformation, the parameter of which is then promoted to a
dynamical field by using a Fadeev-Popov type trick.
In this formulation the bosonic and fermionic theories come as
two different gauge choices of an enlarged gauge symmetry,
and so the equivalence need not apply to off-shell quantities,
such as the effective action, which are not \apriori\ gauge independent.
By contrast, in the approach presented here both theories are
obtained using the {\it same} gauge and so their equivalence is manifest
even off shell. It might be interesting to explore, in the present formulation,
the more general, on-shell, equivalence that becomes available by using
different gauges as is done in ref. \damgaard.  Furthermore,
since the collective field method does not rely on the existence of symmetries
in the original theory, it might potentially be used to generalize duality
to a more general context.

Finally, we remark that our approach opens up a number of interesting
questions.
Since dualization is not an intrinisically two-dimensional procedure,
perhaps it could
be used to provide a prescription for bosonization in higher-dimensional
fermionic field theories. Also, an equivalent understanding of how to
fermionize
bosonic theories could lead to new types of duality transformations
 which could
have wider applications.  We intend to pursue some of these issues
in a future publication.

\bigskip

\centerline{\bf Acknowledgments}

\bigskip

We would like to acknowledge helpful conversations with Poul Damgaard,
Xenia de la Ossa, Oscar Hern\'andez and Rob Myers.
This research was partially funded by N.S.E.R.C.\ of Canada,
les Fonds
F.C.A.R.\ du Qu\'ebec, and the Swiss National Foundation.

\listrefs

\bye